\documentclass[10pt, hyperref]{iopart}
\usepackage{siunitx}
\usepackage{iopams} 
\usepackage{graphicx}
\usepackage{dcolumn}
\usepackage{bm}
\usepackage{braket}
\usepackage{upgreek}
\usepackage{enumitem}
\usepackage[pagebackref]{hyperref}
\hypersetup{
    colorlinks,
    linkcolor={black},
    citecolor={blue},
    urlcolor={blue}
}
\usepackage{xcolor}

\newcommand{\statea}{\ket{6S_{1/2}, F=4, m_F=-4}}
\newcommand{\stateb}{\ket{6S_{1/2}, F=3, m_F=-3}}
\newcommand{\statee}{\ket{6P_{3/2}, F^\prime=4, m_{F^\prime}=-4}}
\newcommand{\Omegatwo}{\tilde{\Omega}}
\newcommand{\Gammatwo}{\tilde{\Gamma}}

\newcommand{\Omegagen}{\tilde{\Omega}_\text{g}}
\newcommand{\deltatwo}{\tilde{\delta}}
\newcommand{\ODtwo}{\tilde{\text{OD}}}

\begin{document}

{\centering

\title{Observation of oscillatory Raman gain associated with two-photon Rabi oscillations of nanofiber-coupled atoms }

\author{Christian Liedl$^1$, Sebastian Pucher$^1$, Philipp Schneeweiss$^1$, Leonid P. Yatsenko$^2$, and Arno Rauschenbeutel$^1$}

\address{$^1$ Department of Physics, Humboldt-Universität zu Berlin, Unter den Linden 6, 10099 Berlin, Germany}
\address{$^2$ Institute of Physics, National Academy of Sciences of Ukraine, prospect Nauki 46, Kyiv-39,
03650, Ukraine} 
\ead{arno.rauschenbeutel@hu-berlin.de}

\begin{abstract}
Quantum emitters with a $\Lambda$-type level structure enable numerous protocols and applications in quantum science and technology. Understanding and controlling their dynamics is, therefore, one of the central research topics in quantum optics. Here, we drive two-photon Rabi oscillations between the two ground states of cesium atoms and observe the associated oscillatory Raman gain and absorption that stems from the atom-mediated coherent photon exchange between the two drive fields. The atoms are efficiently and homogeneously coupled with the probe field by means of a nanofiber-based optical interface. We study the dependence of the two-photon Rabi frequency on the system parameters and observe Autler-Townes splitting in the probe transmission spectrum. Beyond shedding light on the fundamental processes underlying two-photon Rabi oscillations, our method could also be used to investigate (quantum) correlations between the two drive fields as well as the dynamical establishment of electromagnetically induced transparency.
\end{abstract}}

\ioptwocol

\section{Introduction}
Quantum emitters with a $\Lambda$-type level structure play a central role in quantum optics. Suitably driving the two allowed optical transitions, ground-state coherences and populations can be controlled and manipulated. This enables, e.g., stimulated Raman adiabatic passages~\cite{vitanov2001laser, bergmann1998coherent} and coherent population trapping~\cite{arimondo1996coherent}. Moreover, the optical response of an ensemble of $\Lambda$-type emitters can be tailored, leading to, e.g., electromagnetically-induced transparency (EIT)~\cite{fleischhauer2005electromagnetically}, as well as slow and stored light~\cite{hau1999light, kozhekin2000quantum}. 
In all these examples, the driven ensemble closely follows its steady state. The transient dynamics is nevertheless relevant, e.g., to the ultimate efficiency and fidelity of these coherent processes and protocols. 

Two-photon Rabi oscillations between the two ground states of $\Lambda$-type emitters are a prime example of such transient dynamics~\cite{hatanaka1975transient}. In order to observe them, one typically exposes the emitters to a two-photon Rabi pulse with varying duration, followed by a read-out of the ground-state populations \cite{gentile1989experimental, linskens1996two}. This method does, however, not give access to the concomitant coherent dynamics of the driving light fields. Their atom-mediated coherent photon exchange manifests as oscillatory Raman gain and absorption, where gain of one field is accompanied by absorption of the other. 

In order to observe this coherent gain and absorption dynamics, one has to maximize the atom number as well as the atom-light coupling strength. Moreover, the two-photon Rabi frequency has to be well defined, meaning that each atom has to be exposed to the same light intensity. In a typical experimental situation, where an atomic ensemble couples to free-space laser beams, these requirements are challenging to meet simultaneously. To our knowledge, an oscillatory Raman gain associated with two-photon Rabi oscillations has therefore not yet been observed. In this context, coupling atoms to guided light fields in nanophotonic structures may turn out advantageous as it allows for efficient and homogeneous coupling of atomic ensembles~\cite{sheremet2021waveguide}. 

Such a waveguide quantum electrodynamics platform can, for example, be implemented with laser-cooled atoms coupled to optical nanofibers~\cite{nayak2018nanofiber, solano2017optical, nieddu2016optical}. In spite of the close vicinity of the nanofiber surface, ground-state decoherence times on the order of milliseconds have been experimentally demonstrated for nanofiber-coupled atoms~\cite{reitz2013coherence}. Experiments employing a $\Lambda$-type level structure in this setting include, e.g., EIT-based light storage~\cite{sayrin2015storage, gouraud2015demonstration}, Raman cooling~\cite{ostfeldt2017dipole,meng2018near}, generation of single collective excitations~\cite{corzo2019waveguide}, probing of the in-trap atomic motion~\cite{markussen2020measurement}, and non-reciprocal Raman amplification~\cite{pucher2022atomic}. 

Here, we drive two-photon Rabi oscillations of a nanofiber-coupled ensemble of atoms using a nanofiber-guided probe field and a free-space coupling field. We observe an oscillatory Raman gain of the probe, which accompanies the two-photon Rabi oscillations. We experimentally confirm that the two-photon Rabi frequency scales as expected with the probe power and the two-photon detuning. This allows us to infer the atomic coupling strength to the nanofiber-guided mode. Finally, we observe an Autler-Townes splitting in the transmission spectrum of the guided probe field, allowing us to calibrate the Rabi frequency of the coupling laser field.

\section{Method and experimental setup}

\begin{figure}
  \centering
  \includegraphics[width=\columnwidth]{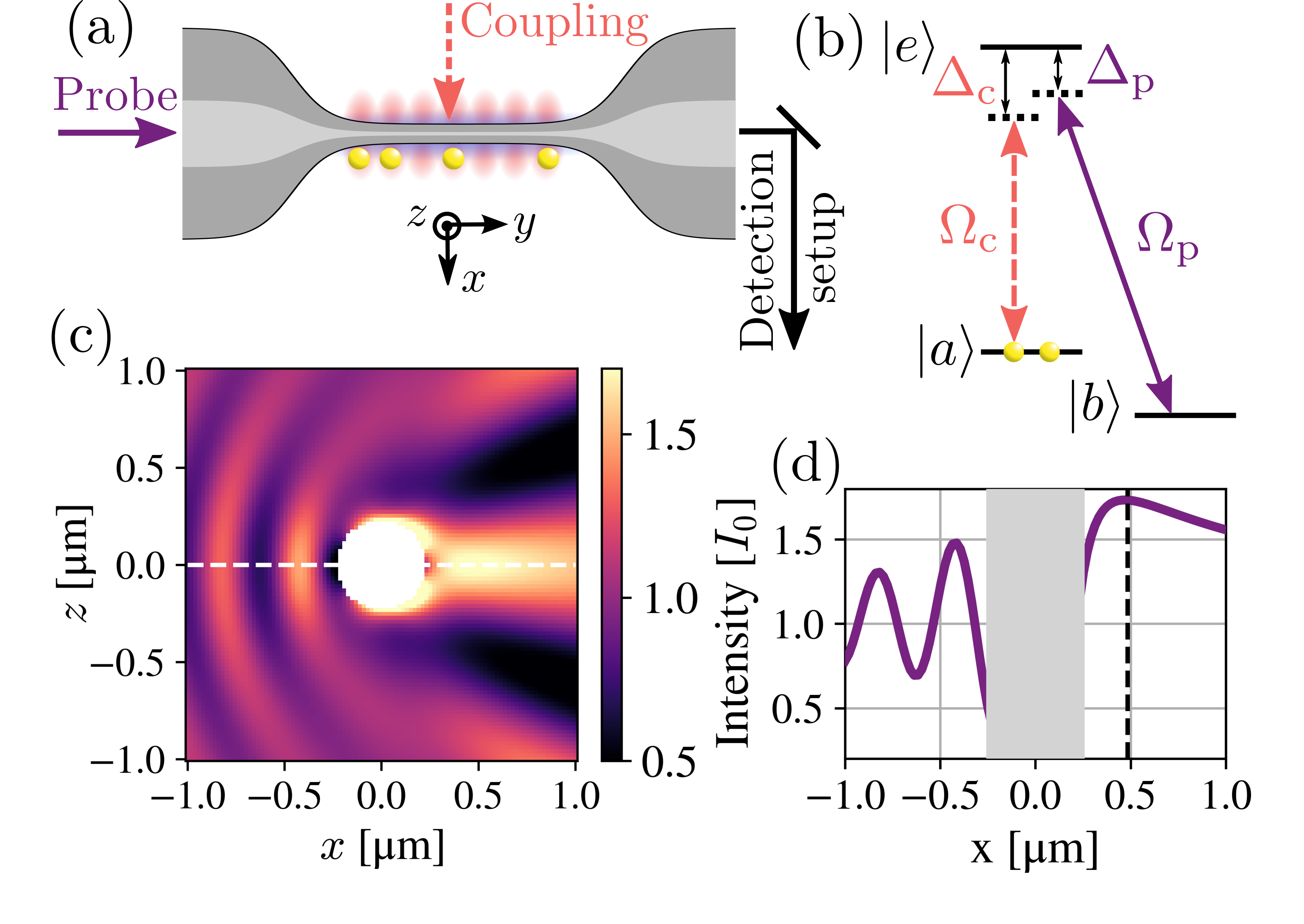}
\caption{(a) Schematic of the experimental setup. Cs atoms (yellow spheres) are optically trapped and interfaced using the evanescent field of the nanofiber waist of a tapered optical fiber. A nanofiber-guided probe laser field and a free-space coupling laser field drive the atoms. We detect the transmitted probe power using two single-photon counting modules (not shown). A magnetic offset field is applied along the $+z$-direction. 
(b) The relevant Cs energy levels form a $\Lambda$ system with two ground states, $\ket{a}$ and $\ket{b}$, and an excited state $\ket{e}$. The coupling- and probe laser fields with Rabi frequencies $\Omega_\text{c}$ and $\Omega_\text{p}$ are detuned by $\Delta_\text{c}$ and $\Delta_\text{p}$ from the excited state, respectively. (c) Intensity distribution in the $x$-$z$ plane of the coupling laser beam in the vicinity of the nanofiber. At large distances from the nanofiber, the coupling light can be approximated as a plane wave with intensity $I_0$, a linear polarization along $z$, and a wave vector pointing into the $+x$-direction. (d) Intensity along the white dashed line shown in panel (c). In front of the nanofiber (left half of the panel), the scattered field features a standing wave pattern due to the reflection off the nanofiber. Behind the nanofiber (right half of the panel), the light is focused, and its intensity is enhanced by a factor of about 1.7 at the position of the atoms, indicated by the black vertical line.}
\label{fig:setup}
\end{figure}

Figure~\ref{fig:setup}(a) schematically shows the core elements of our experimental setup. We optically trap and interface cesium (Cs) atoms using the evanescent field of an optical nanofiber ($\SI{500}{nm}$ nominal diameter) that is implemented as the waist of a tapered optical fiber. We use two nanofiber-guided trapping laser fields to form two diametral arrays of trapping sites along the nanofiber~\cite{vetsch2010optical}. The blue-detuned running wave trapping field has a wavelength of $\SI{760}{nm}$ and a power of $\SI{20.5}{mW}$. The red-detuned, standing wave trapping field has a wavelength of $\SI{1064}{nm}$ and a total power of $\SI{2.4}{mW}$. The minima of the resulting trapping potential are located about $\SI{230}{nm}$ from the nanofiber surface.

We probabilistically load Cs atoms into the trapping potential from a magneto-optical trap using an optical molasses stage~\cite{vetsch2010optical}. Due to the collisional blockade effect, there is at most one atom per trapping site~\cite{schlosser2002collisional}. The atoms on one side of the nanofiber are then further cooled by degenerate Raman cooling using a nanofiber-guided laser field that is near-resonant with the Cs D2 cycling transition~\cite{meng2018near}. Simultaneously, the atoms on the other side are subject to degenerate Raman heating and are thus expelled from the trap. After these steps, we are left with a one-dimensional array of a few hundred atoms in the $\ket{a}=\ket{6S_{1/2}, F=4, m_F=-4}$ state, marking the initial setting for all experiments described below. 

In each run, we determine the optical depth of the trapped ensemble by scanning the fiber-guided probe field over the D2 cycling transition and fitting the resulting transmission spectrum to a saturated Lorentzian absorption profile. We can illuminate the atoms with a free-space coupling laser field propagating in the $+x$-direction. In order to estimate the coupling laser intensity at the position of the atoms, we assume an incident plane wave with intensity $I_0$ that propagates in the $+x$-direction and is linearly polarized along $z$ to compute the intensity distribution of the scattered field around the nanofiber, see Fig.~\ref{fig:setup}(c)~\cite{mitsch2014quantum}. Figure~\ref{fig:setup}(d) shows a cut through the intensity distribution along $x$ for $z=0$. The black dashed line indicates the location of the trapped atoms. Behind the fiber (right half of panel (d)), the intensity of the diffracted coupling laser beam has a maximum close to the position of the trapped atoms. Thus, the latter are exposed to an approximately constant coupling laser intensity even when considering the thermal motion of the atoms in the trap (FWHM of about 100~nm for a temperature of 30~$\upmu$K). Due to the focusing effect of the nanofiber, the atoms are exposed to an intensity that is about 1.7 times larger than $I_0$ at the position of the trap minima, according to our calculation. 

In order to stabilize the atomic population in $\ket{a}$ and suppress spin flips due to spin-motion coupling~\cite{dareau2018observation}, we apply a magnetic offset field in the $+z$-direction. The relevant energy levels form a $\Lambda$ system, see Fig.~\ref{fig:setup}(b). The coupling laser field drives the transition between ground state $\ket{a}$ and excited state $\ket{e}=\statee$, whereas the guided probe field couples the same excited state with the ground state $\ket{b}=\stateb$. The coupling field is $\pi$-polarized, has a Rabi frequency $\Omega_\text{c}$, and is detuned by $\Delta_\text{c}$ from the $\ket{a}\to\ket{e}$ transition. The probe field with Rabi frequency $\Omega_\text{p}$ is phase-locked to the coupling field. It is detuned by $\Delta_p$ from the $\ket{b}\to\ket{e}$ transition and predominantly $\sigma^-$-polarized at the position of the atoms~\cite{mitsch2014quantum}. The excited state $\ket{e}$ has a natural decay rate of $\Gamma=2\pi\times\SI{5.2}{\mega\hertz}$ and decays to $\ket{a}$ and $\ket{b}$ with branching ratios of $7/15$ and $5/12$, respectively. With a probability of $7/60\approx0.1$, $\ket{e}$ decays to $\ket{6S_{1/2}, F=4, m_F=-3}$, such that the three-level system is not closed. However, this does not substantially affect the dynamics on the short timescale we are interested in, which is why we neglect it in the following.

\section{Results and discussion}
\subsection{Autler-Townes splitting}
\begin{figure}[htb]
  \centering
  \includegraphics[width=\columnwidth]{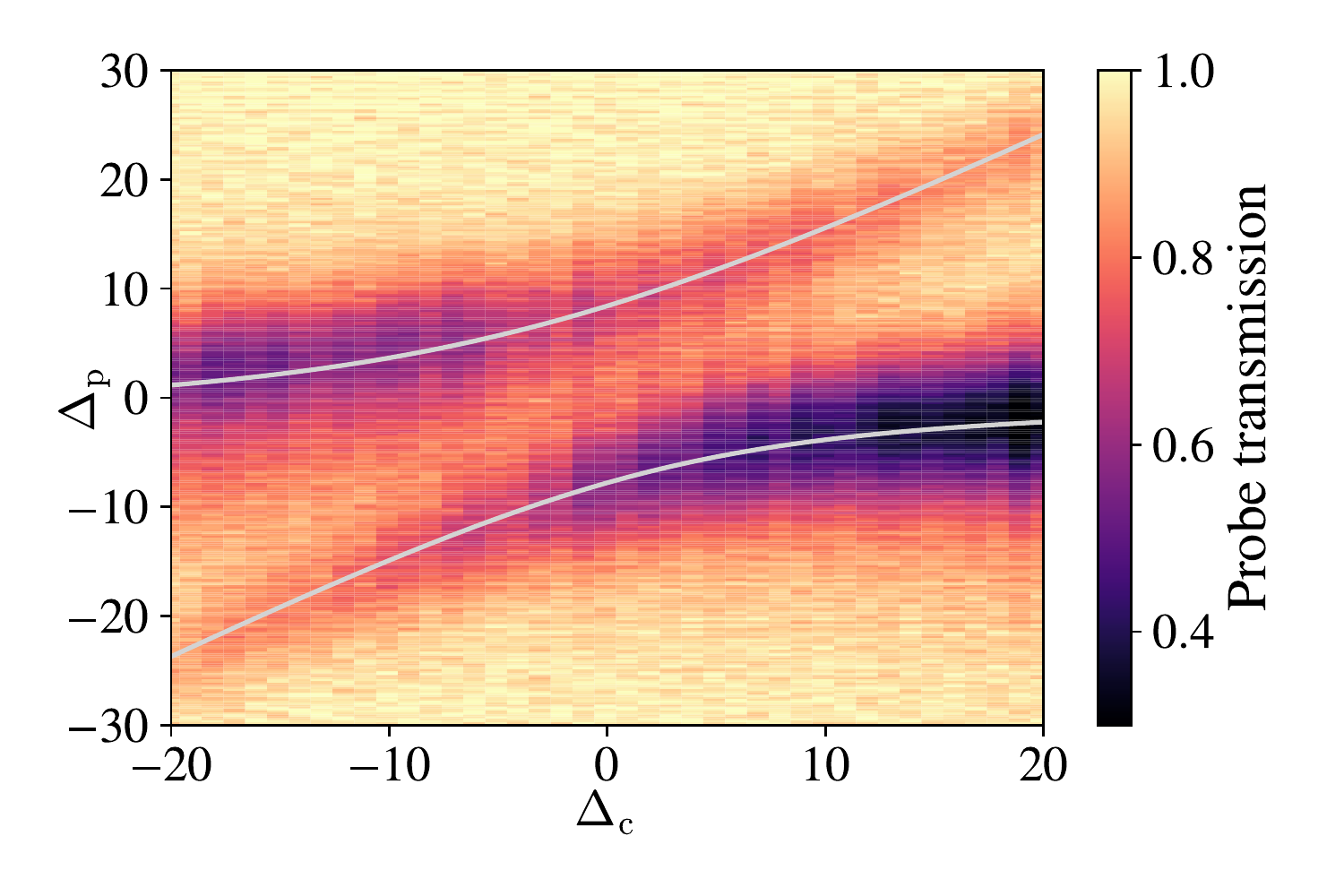}
\caption{Probe transmission spectra for different coupling laser detunings. Two resonances are clearly visible in each spectrum, showing an avoided crossing. By fitting the frequency difference of the two resonances using the prediction based on the Autler-Townes effect (solid gray lines), we find a coupling laser Rabi frequency of $\Omega_\text{c}=\SI{16.1(1)}{\mega\hertz}$. Each spectrum for a given $\Delta_\text{c}$ is averaged over 500 frequency scans of $\Delta_\text{p}$.}

\label{fig:ATspectrum}
\end{figure}

We now turn on the coupling field with a peak intensity of $I_0\approx\SI{25}{mW\per\cm^2}$, which optically pumps all atoms to $\ket{b}$. After $\SI{0.1}{ms}$, we additionally turn on a weak, fiber-guided probe field (saturation parameter of about 0.1) and scan its detuning $\Delta_\text{p}$ over $\SI{60}{\mega\hertz}$ in about $\SI{700}{\micro s}$. This yields a transmission spectrum for a given detuning, $\Delta_\text{c}$, of the coupling laser field. Then, we repeat the measurement for various $\Delta_\text{c}$. The resulting transmission spectrum as a function of $\Delta_\text{p}$ and $\Delta_\text{c}$ is plotted as a colormap in Fig.~\ref{fig:ATspectrum}.
We observe an avoided crossing of two resonances which can be explained by the Autler-Townes effect~\cite{autler1955stark, cohen1996autler, kumar2015autler}: The strong driving of the atoms by the coupling laser field on the $\ket{a}\to\ket{e}$ transition mixes the bare states so that new eigenstates are formed. The energies of those states are split and shifted from $\ket{e}$ by
\begin{equation}
\delta_\text{AT}^{\pm} = \frac{1}{2}\left(\Delta_\text{c}\pm\sqrt{\Omega_\text{c}^2+\Delta_\text{c}^2}\right).
\label{eq:AT_shift}
\end{equation}
We fit the measured spectra to a double Lorentzian line shape and extract the splitting between the two resonance centers for each $\Delta_\text{c}$. We then fit the splitting as a function of $\Delta_\text{c}$ using Eq.~\ref{eq:AT_shift}, yielding a coupling laser Rabi frequency of $\Omega_\text{c} = 2\pi\times\SI{16.1(1)}{\mega\hertz}$. This is in reasonable agreement with the expected $2\pi\times\SI{15.4}{\mega\hertz}$ estimated from the power and diameter of the coupling laser beam used in the experiment and the calculated intensity pattern around the nanofiber. The location of the two resonances assuming the fitted $\Omega_\text{c}$ is in excellent agreement with the data and displayed as the solid gray lines in Fig.~\ref{fig:ATspectrum}. For large $\Delta_\text{c}$, one of the resonances converges to $\Delta_\text{p} \simeq \Delta_\text{c}$. In the following, we will denote the probe detuning from this light-shifted two-photon resonance as $\deltatwo=\Delta_\text{p}-\delta_\text{AT}^+$. 

\subsection{Effective two-level system}
If one chooses large detunings $\Delta_\text{p}$ and $\Delta_\text{c}$, and two-photon detunings ${\deltatwo\approx0}$, such that state $\ket{e}$ is almost unpopulated, then the excited state can be adiabatically eliminated and the $\Lambda$ system can effectively be described by a two-level system~\cite{fleischhauer2005electromagnetically, brion2007adiabatic}. These effective two-level atoms can then undergo coherent two-photon Rabi oscillations between the two ground states $\ket{a}$ and $\ket{b}$, where the two-photon Rabi frequency is given by
\begin{equation}
\Omegatwo=\frac{\Omega_\text{p}\Omega_\text{c}}{2\Delta_\text{c}}~.
\label{eq:Omega_2p}
\end{equation}
The incoherent one-photon scattering of the coupling field
that gives rise to a residual population of the excited state and a subsequent decay can be described by introducing an effective decay rate $\Gammatwo$ for the effective two-level system from $\ket{a}$ to $\ket{b}$
\begin{equation}
\Gammatwo = \frac{1}{8}\frac{|\Omega_\text{c}|^2}{\Delta_\text{c}^2}\Gamma.
\label{eq:Gamma_2p}
\end{equation}
The one-photon scattering of the probe field would correspond to a decay from $\ket{b}$ to $\ket{a}$ in the effective two-level model. However, the one-photon scattering rate of the probe field is small compared to the one-photon scattering rate of the coupling field for our settings, $\Omega_\text{p}^2/\Omega_\text{c}^2 \approx 0.05$, which is why we only take into account the latter in our model. A more detailed description of the adiabatic elimination can be found in the supplementary material. In order to maximize the modulation of the probe transmission caused by Raman gain and absorption, the probe power, and hence, its Rabi frequency should be as small as possible. However, in the absence of technical dephasing, a necessary requirement to see oscillations is $\Omegatwo>\Gammatwo$. Using Eqs.~\ref{eq:Omega_2p} and~\ref{eq:Gamma_2p}, this condition translates to a requirement for the probe Rabi frequency:
\begin{equation}
\Omega_\text{p} > \frac{\Omega_\text{c}\Gamma}{4\Delta_\text{c}}
\label{eq:condition_2p}
\end{equation}
The single-atom dynamics for resonant driving can be described analytically~\cite{torrey1949transient} using the parameters Rabi frequency $\Omegatwo$, decay rate $\Gammatwo$ and decoherence rate $\gamma=\Gammatwo$, see supplementary material.
In principle, fluctuations of the Rabi frequencies due to the thermal motion of the atoms would lead to dephasing, which we neglect here. This approximation is justified a posteriory by the agreement of our model with the experimental data. Another dephasing mechanism arises from the change of the probe power along the atomic ensemble, which can be accurately modeled by consecutively solving the Lindblad master equation and computing the transmission coefficient of each three-level atom~\cite{pucher2022atomic}. We numerically checked that our approach captures the dynamics well in the parameter regime studied here.
Therefore, we simplify the description by assuming that the entire ensemble evolves according to the single-atom dynamics such that the transmission of the probe beam is given by

\begin{equation}
T_\text{probe}(t) = \exp\left[-\ODtwo\frac{\Gammatwo}{\Omegatwo}v(t)\right]
\label{eq:transmission}
\end{equation}
Here, $\ODtwo$ is the optical depth for the probe field on the light-shifted two-photon resonance, and \mbox{$v(t) = 2\Im(\rho_{ba})$}, where $\rho_{ba}$ is the off-diagonal element of the density matrix and $\Im(\ldots)$ denotes the imaginary part. We find that $v(t)$ is given by

\begin{eqnarray}
v(t)\!=\!\frac{\Gammatwo\Omegatwo}{\Gammatwo^2+\Omegatwo^2}\!\left[1\!-\!e^{-\Gammatwo t}\!\left(\!\cos\Omegatwo t\!+\!\frac{2\Gammatwo^2\!+\!\Omegatwo^2}{\Omegatwo \Gammatwo}\!\sin\Omegatwo t\right)\!\right]\!.
\label{eq:sigmay}
\end{eqnarray}

\subsection{Observation of oscillatory Raman gain and absorption}
For the following measurements, we increase the coupling laser detuning to $\Delta_\text{c}\approx2\pi\times\SI{32.5(3)}{MHz}$ and set $\Omega_\text{c}\approx 2\pi\times\SI{28.2}{MHz}$. The probe Rabi frequency can be expressed in terms of the so-called beta factor. It is defined as $\beta=\Gamma_\text{g}/\Gamma$, i.e., the ratio between the atomic emission rate into the guided probe mode $\Gamma_\text{g}$ and the total emission rate $\Gamma$ of the atom. With that, the probe Rabi frequency is
\begin{equation}
\Omega_\text{p} = \sqrt{\frac{5}{12}\frac{4\beta\Gamma P_\text{p}}{\hbar\omega}}~,
\label{eq:Omega_p}
\end{equation}
where $P_\text{p}$ is the input probe power and $\hbar\omega$ the photon energy. In order to increase the coupling strength of the atoms to the guided mode, $\beta$, we increase the red trapping power to $\SI{2.8}{mW}$ during the probing. This shifts the trapping minima closer toward the nanofiber surface and results in a coupling strength of ${\beta\approx0.015}$. 
The probe power of ${\approx\SI{480}{pW}}$ then corresponds to a Rabi frequency of $\Omega_\text{p}=2\pi\times\SI{6.5}{MHz}$.

\begin{figure}
  \centering
  \includegraphics[width=\columnwidth]{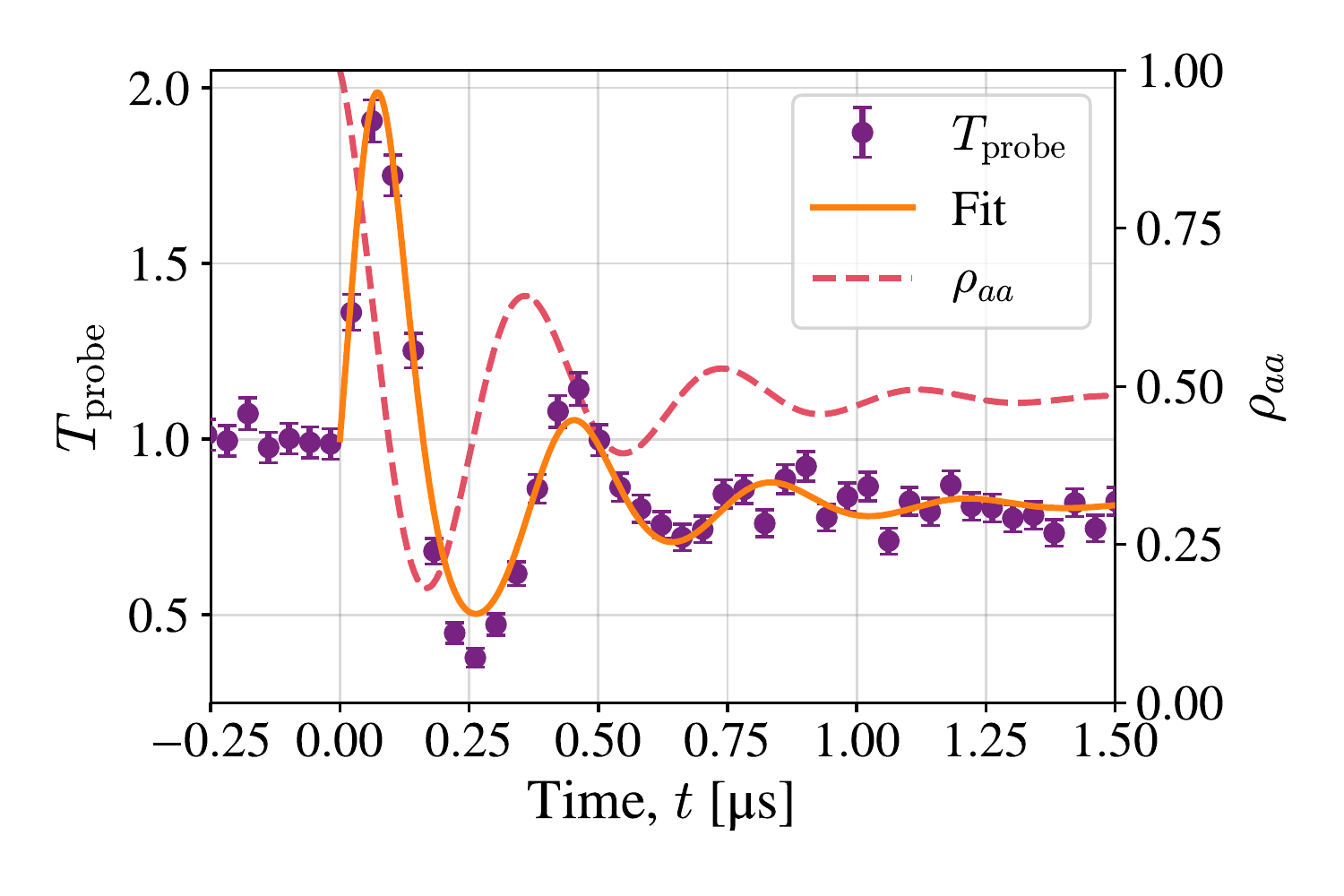}
\caption{Dynamics of the probe transmission coefficient. At time $t=\SI{0}{\micro\second}$, we turn on the coupling laser. The ensemble then undergoes two-photon Rabi oscillations, evident by the oscillatory Raman gain of the probe (purple dots). The error bars are calculated assuming Poissonian counting statistics. We fit the data using Eq.~\ref{eq:transmission} (solid orange line), yielding a two-photon Rabi frequency of $2\pi\times\SI{2.63(3)}{MHz}$, in agreement with an estimate based on independent calibration measurements. Using the fit results, we also infer the average population in $\ket{a}$ (dashed red line). The shown data is averaged over 2300 experimental runs.}
\label{fig:time_trace}
\end{figure}

We plot the measured time-dependent probe transmission for these parameters and ${\approx 700}$ atoms in Fig.~\ref{fig:time_trace}. In the first $\SI{200}{\micro\second}$, only the probe field is on, and we observe unity transmission. This is expected since the probe field does not couple to atoms in the initial state $\ket{a}$. We then turn on the coupling field at time $t=\SI{0}{\micro s}$. 
Since the initial state corresponds to full inversion of the effective two-level system, the transmission of the probe field first experiences Raman gain, and we observe a transmission coefficient of up to ${\approx 2}$. However, during the following coherent dynamics, the transmission oscillates between Raman gain and Raman absorption. After about two full oscillations, the system reaches a steady state. We fit the experimental data in Fig.~\ref{fig:time_trace} using Eq.~\ref{eq:transmission} with $\Omegatwo, \Gammatwo$, and $\ODtwo$ as free fit parameters. The fit (orange solid line) agrees very well with the data for an effective Rabi frequency of $\Omegatwo=2\pi\times\SI{2.63(3)}{MHz}$, in reasonable agreement with the estimated $\Omegatwo=2\pi\times\SI{2.83}{MHz}$, based on independently calibrated parameters. The fit result $\Gammatwo=2\pi\times\SI{516(21)}{kHz}$ is in good agreement with the expected value according to Eq.~\ref{eq:Gamma_2p}, $\Gammatwo=2\pi\times\SI{490}{kHz}$. By assuming that $\Gammatwo=\Gamma\rho_{ee}$, we can infer an average excited state population of $\rho_{ee} = 0.100(1)$.
For the optical density, we obtain $\ODtwo=5.6(2)$. Using these fit results, we also calculate the dynamics of the population in state $\ket{a}$, given by $\rho_{aa}$ (dashed red line in Fig.~\ref{fig:time_trace}). The population dynamics exhibits an oscillation that is $\SI{90}{\degree}$ out of phase with the probe transmission and that reaches a steady-state value of 0.481(4). 
 
\subsection{Scaling of the two-photon Rabi frequency}
\begin{figure}
  \centering
  \includegraphics[width=\columnwidth]{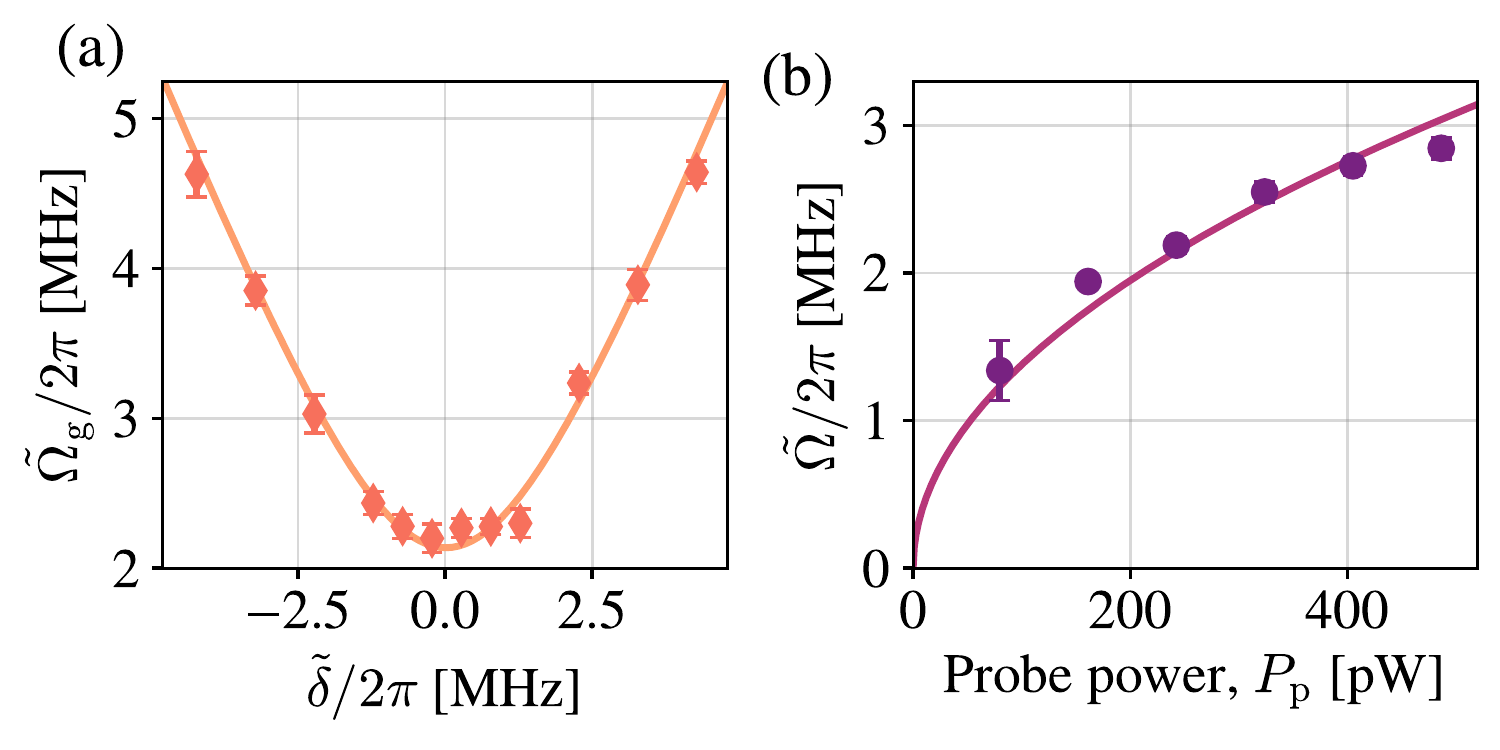}
\caption{Scaling of the two-photon Rabi frequency $\Omegatwo$. (a)~For a two-photon detuning $|\deltatwo|>0$, we observe an increase of the oscillation frequency according to the generalized Rabi frequency $\Omegagen = (\Omegatwo^2+\deltatwo^2)^{1/2}$ (orange diamonds). A fit (orange line) yields $\Omegatwo=2\pi \times \SI{2.13(4)}{\mega\hertz}$. We average over 1000 experimental runs for each $\deltatwo$. (b) Measured values of  $\Omegatwo$ as a function of probe power (purple dots) and fit using a square root function (purple line). We average over 2000 experimental runs for each probe power.}
\label{fig:scalings}
\end{figure}

To gain more insight into the underlying physics of the two-photon Rabi oscillations observed in Fig.~\ref{fig:time_trace}, we repeat the above measurement for different $\deltatwo$ and plot the resulting fitted two-photon Rabi frequency in Fig.~\ref{fig:scalings}(a). These extracted Rabi frequencies can be very well described with the generalized two-photon Rabi frequency, $\Omegagen = (\Omegatwo^2+\deltatwo^2)^{1/2}$. A fit  (orange line) then yields an on-resonance two-photon Rabi frequency $\Omegatwo=2\pi\times\SI{2.12(4)}{MHz}$ for this set of measurements. This is in reasonable agreement with the calculated $2\pi\times\SI{2.0}{MHz}$ for a probe power of ${\approx\SI{240}{pW}}$ used in this measurement. 

In Fig.~\ref{fig:scalings}(b), we show the results of our experimental investigation of the dependence of $\Omegatwo$ on the probe power. We observe that the two-photon Rabi frequency increases with $P_\text{p}$, and we fit the data with a square root function, $A\sqrt{P_\text{p}}$, yielding ${A=2\pi\times\SI{138(3)}{kHz\per\sqrt{pW}}}$. The fit curve (purple line) agrees very well with the data, as expected, given the square root dependence of the Rabi frequency on the power. Using our independent calibrations for $\Omega_\text{c}$ and $\Delta_\text{c}$, we extract $\beta=0.0171(8)$ from the value of $A$, in reasonable agreement with our expectation.
\section{Conclusion and outlook}
In summary, our system features an efficient and homogeneous coupling of the nanofiber-trapped atoms to the guided probe field. In conjunction with a large optical depth of the atomic ensemble on the two-photon resonance, this allowed us to study the oscillatory Raman gain and absorption associated with two-photon Rabi oscillations. In the present work, only the probe field was nanofiber-guided, giving us access to the gain and absorption dynamics of the latter. Launching both probe and coupling field through the nanofiber, as, for example, has been done in~\cite{sayrin2015storage, ostfeldt2017dipole}, would yield the possibility to also measure the correlations between the probe and coupling fields, possibly even with single-photon resolution. Beyond shedding light on the fundamental processes underlying two-photon Rabi oscillations, our method could also be used to investigate the dynamical establishment of electromagnetically induced transparency.

\ack
We would like to thank M. Fleischhauer for stimulating discussions and helpful comments. We acknowledge funding by the Alexander von Humboldt Foundation in the framework of the Alexander von Humboldt Professorship endowed by the Federal Ministry of Education and Research. L.P.Y. is grateful to the Department of Physics of  Humboldt-Universität zu Berlin for his stay as a guest scientist.

\section*{Data availability statement}
The data that support the findings of this study are available from the corresponding author upon reasonable request.
\section*{References}

\bibliographystyle{iopart-num}

\bibliography{bibliography}

\newpage
\onecolumn
\setcounter{equation}{0} 
\setcounter{figure}{0} 
\setcounter{section}{0}
\renewcommand{\theequation}{S\arabic{equation}}
\renewcommand{\thefigure}{S\arabic{figure}}
\renewcommand{\thesection}{S\arabic{section}}

{\centering
\title[Supplementary Material]{Supplementary Material}
}

\section{Lindblad master equation for the three-level system}
Here, we describe the interaction between two classical laser fields with an atomic $\Lambda$-system, comprising two ground states, $\ket{a}=\statea$ and $\ket{b}=\stateb$, as well as an excited state, $\ket{e}=\statee$.
The $\Lambda$-system is driven by a probe laser field and a coupling laser field.
The $\pi$-polarized coupling field with the Rabi frequency $\Omega_c$ is detuned by $\Delta_c$ from the $\ket{a}\rightarrow\ket{e}$  transition. The $\sigma^-$-polarized probe field with the Rabi frequency $\Omega_p$ is detuned by $\Delta_p$ from the $\ket{b}\rightarrow\ket{e}$ transition. In the rotating wave approximation, the Hamiltonian in matrix form is given by~\cite{bergmann1998coherent}

\begin{equation}
\hat\mathcal{H} =     \hbar\left(
\begin{array}{ccc}
0& {\Omega_{p} / 2}   &  0      \\
{\Omega_{p} /2} & - \Delta_p&      {\Omega _{c}/2}  \\
0&   {\Omega_{c}/2}   &  -\Delta_p+\Delta_c      \\
\end{array}
\right), \label{Hamiltonian_RWA}
\end{equation}
where we have chosen the Rabi frequencies to be real.
The dynamics of the density matrix $\hat\rho$ is then determined by the Lindblad master equation \cite{lindblad1976generators} 

\begin{equation}
\frac{d\hat\rho}{dt} = -\frac{i}{\hbar}[\hat\mathcal{H}, \hat\rho]+\sum_{n=1}^2\frac{1}{2}(2\hat c_n\hat\rho\hat c_n^\dagger-\hat\rho \hat c_n^\dagger\hat c_n - \hat c_n^\dagger\hat c_n\hat\rho).
\end{equation}
Here, $[\ldots,\ldots]$ denotes the commutator and $\hat c_n$ are the collapse operators describing spontaneous emission,
\begin{eqnarray}
\nonumber
\hat c_1  = \sqrt{\Gamma_{ea}}\cdot|a\rangle\langle e|, \\
\hat c_2  = \sqrt{\Gamma_{eb}}\cdot|b\rangle\langle e|,
\end{eqnarray}
where $\Gamma_{ea}=(7/15)\Gamma$ and $\Gamma_{ba}=(5/12)\Gamma$ are the population decay rates from the excited state to $\ket{a}$ and $\ket{b}$, respectively, and $\Gamma=2\pi\times\SI{5.2}{MHz}$ is the total decay rate. There are losses from the three-level system with the rate $\Gamma_\text{loss}=(7/60)\Gamma\approx 0.1\Gamma$, which we neglect in the following since they do not significantly alter the dynamics on the short timescale we are interested in.
The differential equations for the density matrix elements read 

\begin{eqnarray}
    \dot\rho_{aa} &=
  - {  i\over 2} \Omega_c  \left(\rho_{ea}-\rho_{ae}\right)+
  \Gamma_{ea}\rho_{ee},\\  
  \dot  \rho_{bb} &=
  - {  i\over 2} \Omega_p  \left(\rho_{eb}-\rho_{be}\right)+
  \Gamma_{eb}\rho_{ee},\\
 \dot\rho_{ab} &=  -
  {  i\over 2}\left( \Omega_c \rho_{eb}- \Omega_p\rho_{ae}\right)  -i\left(\Delta_c-\Delta_p\right)\rho_{ab},\\
 \dot\rho_{ae} &=  -
  {  i\over 2}    \Omega_c  (\rho_{ee}-\rho_{aa})
+{  i\over 2} \rho_{ab} \Omega_p-
 \left({\Gamma\over 2}+i\Delta_c\right) \rho_{ae},\\
 \dot\rho_{be} &= -
  {  i\over 2}    \Omega_p (\rho_{ee}-\rho_{bb})
+{  i\over 2} \rho_{ba} \Omega_c-
 \left({\Gamma\over 2}+i\Delta_p\right) \rho_{be},\\
  \dot\rho_{ee} &=
    {  i\over 2} \Omega_c  \left(\rho_{ea}-\rho_{ae}\right)+
     {  i\over 2}\Omega_p \left(\rho_{eb}-\rho_{be}\right)
  -\Gamma\rho_{ee}.
\end{eqnarray}

\section{Adiabatic elimination and effective two-level system}
Adiabatic elimination of the excited state means that we can neglect the time derivative of the optical coherences and the population of the excited state, i.e. $\dot\rho_{ae}=\dot\rho_{be}=\dot\rho_{ee}=0$. We can then express the matrix elements that involve the excited state in terms of the coherences and populations of the two ground states:

\begin{eqnarray}
\rho_{ae} &=  {-{i\over 2}\Omega_c(\rho_{ee}-\rho_{aa})
+{i\over 2} \rho_{ab} \Omega_p\over
 {\Gamma\over 2}+i\Delta_c},\\
  \rho_{be} &= {-{i\over 2}\Omega_p(\rho_{ee}-\rho_{bb})
+{i\over 2} \rho_{ba} \Omega_c\over
 {\Gamma\over 2}+i\Delta_p} ,\\
    \rho_{ee} &=
    {  i\over 2} {\Omega_c \over \Gamma} \left(\rho_{ea}-\rho_{ae}\right)+
     {  i\over 2} {\Omega_p \over \Gamma} \left(\rho_{eb}-\rho_{be}\right).
\end{eqnarray}
Since we are interested in the dynamics for large detuning and on two-photon resonance, we assume $\Delta_c=\Delta_p=\Delta$. The condition for the validity of adiabatic elimination is that the excited state population $\rho_{ee}$ is very small, i.e. that
\begin{equation}
\Omega_p^2, \Omega_c^2 \ll 4\Delta^2+\Gamma^2.
\end{equation}
If this condition is met, we can eliminate $\rho_{ee}$ in Eqs.~10~--~12 and insert the resulting expressions into the differential equations for the ground-state density matrix elements, Eqs.~4~--~6, which then read

\begin{eqnarray}
  \dot\rho_{aa} &=
  - \Gamma_{a}  \rho_{aa}+ \Gamma_{ba}  \rho_{bb} - {  i\over 2}\Omegatwo(\rho_{ba}-\rho_{ab}),\\
  \dot\rho_{bb} &=
  - \Gamma_{b}  \rho_{bb}+ \Gamma_{ab}  \rho_{aa} + {  i\over 2}\Omegatwo  ( \rho_{ba}-\rho_{ab}),\\
 \dot\rho_{ab} &=  -{  i\over 2}\Omegatwo(\rho_{bb}-\rho_{aa})
  +i \tilde\Delta  \rho_{ab}-\left({\Gamma_c\over 2}+{\Gamma_p\over 2}\right)\rho_{ab}.
\end{eqnarray}
Here, we have introduced the two photon Rabi frequency
\begin{equation}
\Omegatwo = \frac{\Omega_c\Omega_p}{2\Delta}.
\end{equation}
The transition rates to the excited state from state $\ket{a}$ and $\ket{b}$ due to excitation by the coupling and probe fields is given by 
\begin{equation}
\Gamma_{c, p} = \Gamma\frac{\Omega_{c,p}^2}{4\Delta^2},
\end{equation}
respectively. These rates also lead to a relaxation of the coherence $\rho_{ab}$ with rate $\Gamma_c/2+\Gamma_p/2$. The rates $\Gamma_a, \Gamma_b, \Gamma_{ab}$, and $\Gamma_{ba}$ describe the redistribution of population between the two ground states due to spontaneous emission. They are given by the following expressions:
\begin{eqnarray}
\Gamma_{a}&={  1\over 4} {    \Omega_c^2
 \over \Delta ^2 }(\Gamma-  \Gamma_{ea})=\Gamma_c\left(1-{\Gamma_{ea}\over\Gamma}\right)={8\over 15}\Gamma_c,\\
\Gamma_{ba}&={  1\over 4} {    \Omega_p^2
 \over \Delta ^2 }\Gamma_{ea}=\Gamma_p {\Gamma_{ea}\over\Gamma}={7\over 15}\Gamma_p,\\
\Gamma_{b}&={  1\over 4} {    \Omega_p^2
 \over \Delta ^2 }(\Gamma-  \Gamma_{eb})=\Gamma_p\left(1-{\Gamma_{eb}\over\Gamma}\right)={7\over 12}\Gamma_p,\\
\Gamma_{ab}&={  1\over 4} {    \Omega_c^2
 \over \Delta ^2 }\Gamma_{eb}=\Gamma_c {\Gamma_{eb}\over\Gamma}={5\over 12}\Gamma_c.
\end{eqnarray}
Finally, $\tilde\Delta$ denotes the detuning from the light-shifted  two-photon resonance,
\begin{equation}
\tilde{\Delta} = \Delta_p-\Delta_c+\frac{\Omega_c^2-\Omega_p^2}{2\Delta}.
\end{equation}

\section{Simplified effective two-level system}
To describe the experimental data in the main manuscript, we make the following approximations:
\begin{enumerate}[label={\arabic*.}]
   \item We assume that the system is closed, such that $\Gamma_{ea}+\Gamma_{eb}=\Gamma$.
   \item For simplicity, we  assume that the approximate equality $\Gamma_{ea}\approx\Gamma_{eb}$ is exact: $\Gamma_{ea}=\Gamma_{eb}=\Gamma/2$
   \item Since the probe field is much weaker than the control field, we assume that $\Gamma_p=0$.
   \item We assume that we have tuned the laser frequency to the light-shifted two-photon resonance: $\tilde\Delta=0$.
 \end{enumerate}
Under these assumptions, the different rates simplify such that $\Gamma_a = \Gamma_{ab} = \Gamma_c/2$ and $\Gamma_b=\Gamma_{ba}=0$.
The dynamics of the ground-state populations and coherences is then governed by the following set of equations

\begin{eqnarray}
  \dot\rho_{aa} &=
  - \Gammatwo\rho_{aa} - {  i\over 2} \Omegatwo (\rho_{ba}-\rho_{ab} ),\\
  \dot\rho_{bb} &=
  + \Gammatwo  \rho_{aa} + {  i\over 2}\Omegatwo ( \rho_{ba}-\rho_{ab}),\\
 \dot\rho_{ab} &=  -{  i\over 2}\Omegatwo(\rho_{bb}-\rho_{aa})
  -\Gammatwo\rho_{ab}.
\end{eqnarray}
These equations describe a two-level system with ground state $\ket{b}$ and excited state $\ket{a}$, with an effective decay rate from $\ket{a}\rightarrow\ket{b}$, $\Gammatwo$:
\begin{equation}
\Gammatwo = \frac{\Gamma_c}{2} = \frac{\Omega_c^2}{8\Delta^2}\Gamma.
\end{equation}
Here, the coherence $\rho_{ab}$ decays with the same rate as the population, $\Gammatwo$.
This marks an important difference compared to the usual two-level closed system, in which the coherence would decay with a rate of $\Gammatwo/2$. This additional decoherence channel is associated with a transition to the excited state $\ket{e}$ and a subsequent decay to the initial state. This process does not affect the ground-state populations but leads to additional decoherence.
\section{Analytical solution}
Assuming that the system is closed, $\rho_{aa}+\rho_{bb}=1$, we obtain two equations for the inversion, $w=\rho_{aa}-\rho_{bb}=2\rho_{aa}-1$, and the doubled imaginary part of the coherence, $v=2\Im(\rho_{ba})=i(\rho_{ab}-\rho_{ba})$:

\begin{eqnarray}
 \dot v(t) &=  -\Omegatwo w(t)
  -\Gammatwo v(t),\\
  \dot w(t) &=
  - \Gammatwo(w(t)+1)  +\Omegatwo   v(t).
\end{eqnarray}
We are interested in the initial conditions $w(0)=1, v(0)=0$. The solution is given by

\begin{eqnarray}
v(t)&=\frac{\Gammatwo\Omegatwo}{\Gammatwo^2+\Omegatwo^2}\left[1-e^{-\Gammatwo t}\left(\cos\Omegatwo t+\frac{2\Gammatwo^2+\Omegatwo^2}{\Omegatwo \Gammatwo}\sin\Omegatwo t\right)\right],\\
w(t)&=\frac{\Gammatwo^2}{\Gammatwo^2+\Omegatwo^2}\left[-1+e^{-\Gammatwo t}\left(\frac{2\Gammatwo^2+\Omegatwo^2}{\Gammatwo^2}\cos\Omegatwo t-\frac{\Omegatwo}{\Gammatwo}\sin\Omegatwo t\right)\right].
\label{eq:sigmay}
\end{eqnarray}
In the limit of $\Omegatwo\rightarrow 0$, $v$ reaches a steady-state value of $\Omegatwo/\Gammatwo$. The power transmission coefficient $T$ is then determined by the optical density on the light-shifted two-photon resonance, $\tilde{OD}$. Therefore, in general, we can write
\begin{equation}
T(t) = \exp\left[-\tilde{OD}\frac{\Gammatwo}{\Omegatwo}v(t)\right].
\end{equation}

\end{document}